\documentclass[12pt]{article}
\usepackage{amsmath,amsthm,amsfonts,amssymb,amscd}
\usepackage{graphics}
\usepackage[latin1]{inputenc}

\begin{document}

\newcommand{\la}{{\lambda}}
\newcommand{\ro}{{\rho}}
\newcommand{\po}{{\partial}}
\newcommand{\ov}{\overline}
\newcommand{\re}{{\mathbb{R}}}
\newcommand{\nb}{{\mathbb{N}}}
\newcommand{\Z}{{\mathbb{Z}}}
\newcommand{\Uc}{{\mathcal U}}
\newcommand{\gc}{{\mathcal G}}
\newcommand{\hc}{{\mathcal M}}
\newcommand{\fc}{{\mathcal F}}
\newcommand{\dc}{{\mathcal D}}
\newcommand{\al}{{\alpha}}
\newcommand{\vr}{{\varphi}}
\newcommand{\om}{{\omega}}
\newcommand{\La}{{\Lambda}}
\newcommand{\be}{{\beta}}
\newcommand{\te}{{\theta}}
\newcommand{\Om}{{\Omega}}
\newcommand{\ve}{{\varepsilon}}
\newcommand{\ga}{{\gamma}}
\newcommand{\Ga}{{\Gamma}}
\newcommand{\zb}{{\mathbb{Z}}}
\def\sen{\operatorname{sen}}
\def\Ker{\operatorname{{Ker}}}
\newcommand{\bc}{{\mathcal B}}
\newcommand{\rc}{{\mathcal R}}

\centerline{\large\bf QUANTIFICATION OF ELECTRICAL CHARGES IN}

\centerline{\large\bf KALUZA KLEIN THEORY WITHOUT}

\centerline{\large\bf STRINGS OF DIRAC}

\thispagestyle{empty}

\vskip .3in

\centerline{{\sc Paulo da Costa Moreira}}

\centerline{Comissão Nacional de Energia Nuclear (CNEN), Brazil}

\vskip .1in

\centerline{Adress of contact : Av. N. S. de  Copacabana 960/303}

\centerline{Rio de Janeiro - RJ 22060-000 - Brazil}

\vskip .1in

\centerline{Personal e-mail: moreira@sintrasef.org.br}

\centerline{prjcm@cnen.gov.br}

\vskip .3in

\begin{abstract}
It is proposed a formalism of quantification of the electric charges in the Kaluza Klein theory of five dimensions and a explanation of the cause of the variation of the electromagnetic fine-structure constant in cosmological times.There is a formalism that eliminates the problem that impeached the KK theory  to complete the unification of the electromagnetic and the gravitational fields. The complete unification of these two fields is obtained. The radius of compactification is determined as  70,238  lengths of Planck.

\end{abstract}

\noindent
{\bf PACS:} 04.50.+h, 11.25.Mj

\medskip

\noindent
{\bf KEY WORDS:} Kaluza Klein Theory, Compactification.

\section{Alternative Kaluza Klein theory unifying the eletromagnetic and gravitational fields.}

Consider a space time with the variables $x^0, x^1, x^2, x^3, x^4$, where $x^0$ is the time,  $(x^1, x^2, x^3)$ forms the normal space,  and $x^4$ is a compact almost circular  variable where $0 \le x^4 \le  2\pi r_k$,  approximately, and $r_k$ is the radius of compactification of Klein.

We can do:
\begin{equation}
     u^i = dx^i/ds,\quad   i = 0, 1, 2, 3, 4. \tag{1}
\end{equation}
where
\begin{equation}
    ds^2 = g_{kl}' dx^kdx^l,\quad   k,l = 0, 1, 2, 3, 4.\tag{2}
\end{equation}
and  $g_{kl}'$ is the metric tensor of the space-time of five dimensions. Our convention will be $g_{\al\be}$   to  normal space-time   $\al,\be =   0, 1, 2, 3$. The use of  $R_{ik}', R_{\al\be}$, et coetera will do the same convention.   

We know, naturally the relativistic inertial displacement on the geodesic:
\begin{equation}
    du^i/ds =  -\{{}_k{}^i {}_ l\}' u^ku^l,\quad  i,k, l = 0, 1, 2, 3, 4.\tag{3}
\end{equation}
where $\{{}_k{}^i {}_ l\}$ is the Christoffel symbols. But using the $\{\dots\}'$ to express the Christoffel symbols of the penta-space-time, and using $\{\dots\}$ to the case of the normal space-time. 

Naturally, using  $\al,\be,\ga = 0, 1, 2, 3$, we can do from  (3): 
\begin{equation} 
 du^\al/ds  =  - [\{{}_\be{}^\al {}_\ga\}'u^\be u^\ga + \{{}_\be{}^\al {}_4\}' u^\be u^4 + \{{}_4{}^\al {}_\ga\}' u^4u^\ga +  \{{}_4{}^\al {}_4\}'] u^4u^4 \tag{4}
\end{equation}

\vskip .1in

Electing:

\vskip .1in

 $x_4\equiv$  the length of the arch of the right section of the cylinder of the Universe, 

\vskip .1in

\noindent
or:

\vskip .1in

\noindent
the  coordinate  $x_4$   is defined as how:\,\, if
$$
   dx^0 = dx^1 = dx^2 = dx^3 = 0,
$$
$$
ds = dx_4
$$
therefore
$$
g_{44}'\equiv 1
$$

This choice of the coordinate $x_4$  is very important, because it eliminates the  problem that impeached the Kaluza Klein theory of to do a complete unification between the gravitational field and the electromagnetic field.This is the first time what it is obtained a complete unification of the electromagnetic and the gravitational fields in the Kaluza Klein theory of five dimensions. The tensorial liberty of to choice anyone coordinate is broke in the case of  fifth coordinate. But the experimental requirements of the relativity demands the liberty of choice of the four coodinates of the normal space-time so only. The liberty of choice of $x^4$  is not necessary to establish the relativity of the coordinates $x_0, x_1, x_2, x_3$. This relativity is verified in the experiments of the physic , but demands  not the liberty of choice of  $x_4$. By other side, if we elect the fifth coordinate as we did, we obtain the
quantification of the electrical charges which is a experimental fact of the Nature.

Now, we consider a $\rc^4$  space where there is a 
$$
g^{\al\be} = g^{'\al\be};\quad \al,\be = 0,1,2,3.
$$
Naturally we can define $g_{\al\be}$   the inverse matrix of  $g^{\al\be}$  as:
$$
g_{\al\mu} g^{\mu\be}=\delta_\al^\be.
$$
And without to think in the electromagnetical field (or as it is defined in the classical Kaluza Klein Theory) we will consider the definition of a mathematical function:
$$
A_\be=g_{4\be}'=g_{\be4}'
$$

Naturally we can define:
$$
A^\mu=g^{\mu\be}A_\be
$$
The unique case that preserves the necessary relation: 
$$
g^{'\al\be}g_{\be\mu}'=\delta_\mu^\al
$$
is the case:
\begin{equation}
g_{4\be}'=g_{\be4}'=A_\be;\,\, g_{44}'=1;\,\, g_{\al\be}'=g_{\al\be}+A_\al A_\be;\,\, g^{'\al\be}=g^{\al\be};\,\, g^{'4\be}=g^{'\be4}=-A^\be \tag{5}
\end{equation}
and
$$
g^{44}=1+A^\be A_\be
$$
to $\al,\be=0,1,2,3$.

Considering that there is a symmetry  (cylindrical)  that needs be explained for the cosmologists in the future:
\begin{equation}
\po g_{kl}'/\po x^4 \equiv 0,\quad k,l=0,1,2,3,4\tag{6}
\end{equation}
and  choicing the ``Gauge" (with ``$\dots$", because $A_\be$  is only a mathematical  object here)  :
$$
A_\be u^\be(t)=0
$$
All these hypothesis and the  equation   (4) will result in:
\begin{equation}
du^\al/ds=-  \{{}_\be{}^\al {}_\ga\} u^\be u^\ga+ F^\al_\be u^\be u^4;\quad \al,\be,\ga=0,1,2,3.\tag{7}
\end{equation}
where  we define:
\begin{equation}
F_{\be\ga}=\po A_\ga/\po x^\be-\po A_\be/\po x^\ga.\tag{8}
\end{equation}

\vskip .1in

\noindent
{\bf Demonstration of  (7):}

Repeating  (4)  for  better  clarity:
$$
du^\al/ds=-[\{{}_\be{}^\al {}_\ga\}' u^\be u^\ga+
\{{}_\be{}^\al {}_4\}'u^\be u^4+ \{{}_4{}^\al {}_\ga\}'u^4 u^\ga+ \{{}_4{}^\al {}_4\}']u^4 u^4.
$$
But from (6):  
$$
\po g_{kl}'/\po x^4\equiv 0,\quad k,l=0,1,2,3,4\text{ and } g_{44}=1.
$$
We can conclude  that  $\{{}_4{}^\al {}_4\}' = 0$.

By other side, remembering  from (5) the relations about the metric tensors
$$
\{{}_\be{}^\al {}_4\}'u^\be u^4= \{{}_4{}^\al {}_\ga\}'u^4 u^\ga=(1/2)[g^{\al\mu}\{\mu,\be4\}'-A^\al\{4,\be 4\}']=(1/2)g^{\al\mu}\{\mu,\be4\}'.
$$
Naturally  $\{4,\be 4\} = 0$ because:\,\,   (6)\,\,  $\po g_{kl}'/\po x^4\equiv 0$.

\vskip .1in

Therefore:
$$
 \{{}_\be{}^\al {}_4\}'u^\be u^4+ \{{}_4{}^\al {}_\ga\}'u^4 u^\ga= 2g^{\al\mu}\{\mu,\be4\}'u^\be u^4 = g^{\al\mu}\{\po_\be g_{\mu 4}'+\po_4 g_{\mu\be}'-\po_\mu g_{\be4}'] u^\be u^4.
$$

Considering from (5) $g_{\be4}' = A_\be$   and from (6): $\po g_{kl}'/\po x^4\equiv 0$

Naturally:
\begin{align*}
 \{{}_\be{}^\al {}_4\}'u^\be u^4+ \{{}_4{}^\al {}_\ga\}'u^4 u^\ga &=  2g^{\al\mu}\{\mu,\be4\} u^\be u^4=g^{\al\mu} [ \po_\be g_{\mu4}'-\po_\mu g_{\be 4}'] u^\be u^4 \\
&= g^{\al\mu}[\po_\be A_\mu-\po_\mu A_\be] u^\be u^4 
\end{align*}
therefore:
$$
 \{{}_\be{}^\al {}_4\}'u^\be u^4+ \{{}_4{}^\al {}_\ga\}'u^4 u^\ga = -F^\al_\be u^\be u^4.
$$
We can reduce now (4) to the form:
$$
du^\al/ds=- \{{}_\be{}^\al {}_\ga\}'u^\be u^\ga+F^\al_\be u^\be u^4.
$$
But  
$$
 \{{}_\be{}^\al {}_\ga\}'=(1/2)[g^{\al\mu}\{\mu,\be\ga\}'-A^\al\{4,\be\ga\}']=(1/2)[g^{\al\mu}\{\mu,\be\ga\}'-(1/2)A^\al\{4,\be\ga\}'].
$$
Considering the relations of (5) and (6):  $\{ 4, \be\ga\}'=\po_\be A_\ga+\po_\ga A_\be$.
Therefore
$$
 \{{}_\be{}^\al {}_\ga\}'=(1/2)[2g^{\al\mu}\{\mu,\be\ga\}'-A^\al(\po_\be A_\ga+\po_\ga A_\be)].
$$

By other side as   $g_{\al\be}'=g_{\al\be}+A_\al A_\be$,   
\begin{align*}
 \{{}_\be{}^\al {}_\ga\}'=(1/2)[2g^{\al\mu}\{\mu,\be\ga\} &+ A^\al\po_\ga A_\be+A_\be g^{\al\mu}\po_\ga A_\mu+A^\al\po_\be A_\ga \\
& +A_\ga g^{\al\mu}\po_\be A_\mu-g^{\al\mu}A_\be\po_\mu A_\ga-g^{\al\mu} A_\ga\po_\mu A_\be-A^\al\po_\be A_\ga-\po_\ga A_\be].
\end{align*}

Doing the cancellations:
$$
 \{{}_\be{}^\al {}_\ga\}'=(1/2)[2g^{\al\mu}\{\mu,\be\ga\}+ A_\be g^{\al\mu}\po_\ga A_\mu+A_\ga g^{\al\mu}\po_\be A_\mu-g^{\al\mu}A_\be\po_\mu A_\ga-g^{\al\mu}A_\ga\po_\mu A_\be]
$$
therefore:
$$
\{{}_\be{}^\al {}_\ga\}'u^\be u^\ga= \{{}_\be{}^\al {}_\ga\} u^\be u^\ga+(1/2)[A_\be g^{\al\mu}\po_\ga A_\mu+A_\ga g^{\al\mu}\po_\be A_\mu-g^{\al\mu} A_\be\po_\mu A_\ga- g^{\al\mu} A_\ga\po_\mu A_\be] u^\be u^\ga.
$$
But if  $A_\be u^\be(t)=0$  (the ``gauge") we  obtain:
$$
 \{{}_\be{}^\al {}_\ga\}'u^\be u^\ga= \{{}_\be{}^\al {}_\ga\} u^\be u^\ga
$$
and returning to:
$$
du^\al/ds=- \{{}_\be{}^\al {}_\ga\}'u^\be u^\ga+F^\al_\be u^\be u^4
$$
we can conclude the demonstration of (7):
$$
du^\al/ds=- \{{}_\be{}^\al {}_\ga\} u^\be u^\ga+F^\al_\be u^\be u^4.
$$
Q.E.D.

\vskip .2in

Now, we can stop and to think. This kind of mathematical object is similar to a {\bf complete} unification of the gravitational field and the electrical field (without the $g_{44} = - \phi$  to disturb all). The unification  pretended for Kaluza and Klein and never completed, it is here, now. The magic was the definition of the variable:

  $x^4\equiv $ the length of the arch of the right section of the cylinder of the Universe.

That produces:
$$
g_{44}\equiv 1.
$$

\section{Quantification  of  electric charge.}

In the electromagnetism without gravitational field
$$
(mcdu^\al/ds)_{\text{eletromagnetic}} = (q/c) F^\al_\be u^\be.
$$

	It is obvious that:
\begin{equation}
q=cp^4=mc^2 u^4\tag{9}
\end{equation}
does the function of the charge of the particle. But the coordinate $x^4$  is compacted and it is a (almost) ``circle" of  $2\pi r_k$  of extension. Obvious  the wave-particle will have a wave length:
\begin{equation}
\la=2\pi r_k/n,\quad n\in\nb \tag{10}
\end{equation}
and  
\begin{equation}
p^4=h/\la=n h/2\pi r_k.\tag{11}
\end{equation}

Considering that the charge of the particle is:
\begin{equation}
q=cp^4=n\cdot (\hbar c/r_k).\tag{12}
\end{equation}

We conclude that the electric charge is quantified, with a unitary quantum  $e'$, where
\begin{equation}
e'=\hbar c/r_k.\tag{13}
\end{equation}

We know that the quantum of  electric charge $e$ is $1/3$ of the charge of the electron  $e$. Therefore:
\begin{equation}
e=3e'=3\hbar c/r_k.\tag{14}
\end{equation}
It is obvious that it defines the dimensionality of  $q, A_\mu , F^\be_\al$:
\begin{equation}
 [q] =  ML^2/T^2 ,\,\, [ A_\mu] = 1,\,\,  [ qA_\mu] = ML^2/T^2,\,\,  [ F_\be^\al] =  L^{-1},\,\, [qF_\be^\al] =  ML/T^2.\tag{15}
\end{equation}

\section{Calculating the permissivity.}

Using the definition of the Tensor of Riemann  $R_{kj}$  and remembering (5) and (8) it is possible to prove:
\begin{equation}
R_{\be4}'=- D_\al F_\be^\al/2;\quad \al,\be=0,1,2,3.\tag{16}
\end{equation}
Where  $D_\al$  is the covariant derivative.

\vskip .2in

\noindent
{\bf Demonstration:}
Using the definition of the Tensor of  Riemann of second order:
$$
R_{\be4}'=\po_m \{{}_\be{}^m {}_4\}'- \po_4 \{{}_\be{}^m {}_m\}'  + \{{}_\be{}^l {}_4\}' \{{}_l{}^m {}_m\}' - \{{}_\be{}^l {}_m\}'  
\{{}_l{}^m {}_4\}'.
$$
Where  $\be=0,1,2,3$; $m,l=0,1,2,3,4$ and $\{\dots\}'$ denotes the symbol of Christoffel in the $\rc^5$.

But
$$
\po_4 \{{}_\be{}^m {}_m\}' =0;\,\, \{{}_\be{}^m {}_4\}' =- F^m_\be/2;\,\, \{{}_\be{}^l {}_4\}'=- F^l_\be/2; \text{ and } \{{}_l{}^m {}_4\}'=- F^m_l/2.
$$

Doing these substitutions:
$$
R_{\be4}'=-[\po_mF_\be^m+F_\be^l \{{}_l{}^m {}_m\}'- \{{}_\be{}^l {}_m\}'F_l^m]/2 \equiv -D_\al F_\be^\al/2.
$$
It demonstrates the equation (16).

Q.E.D.

\bigskip

But
$$
D_\al F^{\be\al}=j^\be/\ve_0 c\Rightarrow D^\al F_\al^\be= j^\be/\ve_0 c\Rightarrow D^\al F_{\be\al}=j_\be/\ve_0 c\Rightarrow D^\al F_{\al\be}=-j_\be/\ve_0 c
$$

\begin{equation}
D_\al F^\al_\be=-j_\be/\ve_0 c;\quad \al,\be=0,1,2,3. \tag{17}
\end{equation}
Where  $j_\be$ is the electric current and $\ve_0$  is the permissivity.

By other side, another form of the equation of Einstein of gravitation is:
\begin{equation}
R_\al^\be=(8\pi G/c^4)[T_\al^\be-\delta_\al^\be T/(N-2)];\quad \al,\be  = 0, 1, 2, 3 , \text{ or } \al,\be = 0, 1, 2, 3, 4; \tag{18}
\end{equation}
 where $N$ is the number of dimensions of  the Kaluza Klein.  

Where  $T_\al^\be$  is the tensor of  momentum-energy and $T$ is its scalar contraction , and $N$ is the number of dimensions of the considered  theory (Kaluza Klein or traditional).

Naturally:
\begin{equation}
R_\al^{'4}=(8\pi G/c^4) T_\al^{'4},\quad \al =0,1,2,3.\tag{19}
\end{equation}
Consider a special case where the space is almost cylindrical and the charge and the mass of the body  of probe is little:

We will use  $g_{\al\be}'$, $R_{\al\be}'$, et coetera, to impeach the confusion with the traditional case of four dimensions.
\begin{equation}
g_{00}' =1,\quad g_{ii}'=-1,\quad i =1,2,3, \text{ and in our theory } g_{44}'=1;\tag{20}
\end{equation}
by other side:
\begin{align*}
g_{ij}' &= 0 \text{ to } i,j=1,2,3 \text{ and } i\cong j; \\
g_{i4}' &= g_{4i}' = A_i'; \\
g_{0i}' &=  g_{i0}'= 0 \text{ and } g_{04}' = g_{40}' = A_0'.
\end{align*}

Therefore:
\begin{equation}
T_{\al4}'=g_{44} T_\al^{'4}+ g_{4\be} T_\al^{'\be};\quad \al,\be = 0,1,2,3,\tag{21}
\end{equation}
considering (5)
\begin{equation}
T_{i 4}'= T_i^{'4}+A_\be T_\al^{'\be}; \quad \al,\be=0,1,2,3. \tag{22}
\end{equation}

In our special case the body of prove is small (as a baseball ball, for example) and its gravitational field is negligible. If his charge is the minimum ( a third part of the charge of the electron) the electromagnetic field - in  almost all inner volume of the ball - will be negligible and therefore:
\begin{equation}
T_{i4}'\cong T_i^{'4}.\tag{23}
\end{equation}

Considering  (19), it  conduces to:
\begin{equation}
R_{i4}'=8\pi G T_{i4}'/ c^4\cong 8\pi G T_i^{'4}/c^4 . \tag{24}
\end{equation}

By other side:
\begin{equation}
T_\al^{'\be}=(p+\ve)u_\al u^\be+p\delta_\al^\be.\tag{25}
\end{equation}
Where  $p$  the pressure  and  $\ve$  is the density of the energy.

Naturally:
\begin{equation}
T_i^{'4}=(p+\ve) u_i u^4.\tag{26}
\end{equation}

Consider now that the body source of the gravitational  field is naturally little and it is solid too, and the pressure is negligible.
\begin{equation}
T_i^{'4}=mc^2 u_i u^4/V.\tag{27}
\end{equation}
Where  $V$  is the volume of the body  which is the source of the gravitational field  and $m$ is the mass.

Remembering (9), (23) and (27),  we obtain:
\begin{equation}
j_i=qcu_i/V=mc^3 u_i u^4/V = cT_i^4 \cong cT_{i4}', \quad i=1,2,3.\tag{28}
\end{equation}

Considering (16), (17),  and (28)
\begin{equation}
R_{i4}' = j_i/2\ve_0 c \cong T_{i4}'/2\ve_0. \tag{29}
\end{equation}

But, remembering (24), we conclude:
\begin{equation}
\ve_0=c^4/16 \pi G. \tag{30}
\end{equation}

This is the value of the permissivity in this theory.

\section{Calculating the radius of Klein (of the compactification).}

The constant of fine-structure $a$  is:
\begin{equation}
a=e^2/4\pi \ve_0 \hbar c . \tag{31}
\end{equation}

 If we put  (12) and (30) in (31):
\begin{equation}
\al=36 G\hbar/c^3 r^2_k . \tag{32}
\end{equation}
\begin{equation}
\al=36 l^2_{\text{Planck}}/r_k^2.\tag{33}
\end{equation}
Where  $l_{\text{Planck}}$  is the length of Planck.

It conduces to the calculation of the radius of compactification of the Universe (radius de Klein):
\begin{equation}
r_k=70,238 \,\, l_{\text{Planck}}. \tag{34}
\end{equation}

\section{The signal of  $g_{44}$}

The signal of   $g_{44}$  is important. Our preference  was the positive signal. It  causes that in a case without gravity:
\begin{equation}
E^2/c^2  +  (p^4)^2 - ( p_x)^2 - (p_y)^2 - (p_z)^2 =  m^2 c^2.\tag{35}
\end{equation}
But there is a ``effective mass", considering the usual formula:
\begin{equation}
E^2/c^2   - ( p_x)^2 - (p_y)^2 - (p_z)^2 =  m_{eff}^2  c^2.\tag{36}
\end{equation}
Where  $m_{eff}$   is the mass  normally attributed to the particles  and, naturally:
\begin{equation}
m = (m_{eff}^2  + (p^4 )^2 /c^2 )^{1/2} = ( m_{eff}^2  + q^2 /c^4 )^{1/2} =  (m_{eff}^2  + 
 n^2\,  \hbar^2 /c^2r_k^2)^{1/2}.\tag{37}
\end{equation}
The effective mass is the mass that we observe in the particle.

If  the signal of  $g_{44}$  would be  negative  and  $g_{44} = -1 $
\begin{equation}
E^2/c^2  -   (p^4)^2 - ( p_x)^2 - (p_y)^2 - (p_z) =  m^2 c^2\tag{38}
\end{equation}
and
\begin{equation}
m = ( m_{eff}^2  - (p_4)^2 /c^2 )^{1/2} = (m_{eff}^2  - q^2 /c^4)^{1/2} =  ( m_{eff}^2  -  
 n^2\, \hbar^2 /c^2r_k^2)^{1/2}.\tag{39}
\end{equation}

It  would  cause that the real masses of the charged particles (that we know) would be imaginary (without  physical sense). Therefore, the fifth coordinate is a ``kind-time" and no a ``kind-space" coordinate. Another interpretation is to consider that the effective mass would be excessively  big  (as in the hierarchy   problem) if the fifth coordinate would be ``kind-space". Therefore, the fifth coordinate must be ``kind-time".  

\section{The   variation of the fine constant structure in the cosmological times.}

If  the radius of compactification suffers a little change during the cosmological times it can explain the variation of the fine-structure constant [1] of :
  \begin{equation}
\Delta\al/\al\Delta t = (-0,2\pm 0,8) \times 10^{-17} \text{ year}^{-1}\tag{40}
\end{equation}
mentioned for Srianand [2] et al. 

There is a lot of cosmological models and it is necessary  to investigate the variation of  the fine structure constant in all cases. I invite the cosmologists to do this  job.

\section{The  question of the Experimental Test of the Theory of  Quantum Strings}

We knows that the strings models and Kaluza Klein models  converges almost. If this theory of quantum strings is true, the ``circle of Klein" (of the fifth coordinate) needs to be a natural number of  the quantum strings.

Consider that :
\begin{equation}
  l_{\text{string}}  =  \eta\,  l_{\text{Plank}} \tag{41}
\end{equation}
where  $l_{\text{string}}$  is the length of the quantum string and   $\eta$  is a no dimensional constant.

Naturally:
\begin{equation}
2\pi  r_k = N \, \eta\,  l_{\text{Plank}}\tag{42}
\end{equation}

If  we  use  (33):
\begin{equation}
  \alpha =   36\,\, l^2_{\text{Planck}} / r^2_k =  144 \pi^2/ N^2\eta^2 \tag{44}
\end{equation}

The fine structure constant would suffer  quantum jumps during its cosmological variation. In a jump of  $N + 1$ to $N$  would ocurr:
\begin{equation}
 \Delta \alpha/\alpha   =    - (2N + 1)/ N^2(N +1)^2 \tag{45}
\end{equation}

This result is compatible with a variation of  $\sim 10^{-7}$ (in $10^{10}$ years) to $\eta\sim 1$, and consequently (see (34) and (42)) a value of  $N \sim 200$. The model created here and the theory of quantum strings are compatible. But it is necessary better measurements to  confirm the variation of the fine structure constant and to verify if the variation is continuous or for quantum jumps. If it is continuous the theory of quantum strings is false. If it is by quantum jumps this theory is true. Finally we have a manner of to test~it.

\section{Conclusions}

The Kaluza Klein of  five dimensions conduces to the quantification of the charges (without string of Dirac) and creates a probable way of explanation  to the variation of  the fine structure constant.It is possible to do a complete unification between the electromagnetic and gravitational fields, sacrificing the liberty of choice the fifth compact coordinate, without to change the relativity in the four space-time and  naturally doing maintenance of the liberty of to choice of the four coordinates of the normal space-time. It is possible to test, for astronomy, if the quantum strings are true or not.

\vskip .4in 

\noindent
{\bf  REFERENCES}

\begin{itemize}

\item[{[1]}]  Mohr, P.J., Taylor, B.N. {\it Rev. Mod. Phys\/}. {\bf 72}, 351--495 (2000).

\item[{[2]}] arXiv: astro-ph/0402177,  8 Feb. (2004).

\end{itemize}

\end{document}